\documentclass[%
 reprint,
 amsmath,amssymb,
 aps,
pra,
]{revtex4-1}

\usepackage{graphicx}
\usepackage{dcolumn}
\usepackage{bm}

\begin{document}

\preprint{APS/123-QED}

\title{Cryptanalysis of the arbitrated quantum signature protocols}

\author{Fei Gao}
  \email{gaofei\_bupt@hotmail.com}
 \altaffiliation[Also at ]{State Key Laboratory of Integrated Service Networks, Xidian University, Xi'an 710071, China.}
\author{Su-Juan Qin}
\author{Fen-Zhuo Guo}
\author{Qiao-Yan Wen}
\affiliation{
 State Key Laboratory of Networking and Switching Technology, Beijing University of Posts and Telecommunications, Beijing, 100876, China
}

\date{\today}

\begin{abstract}
As a new model for signing quantum message, arbitrated quantum signature (AQS) has recently received a lot of attention. In this paper we study the cryptanalysis of previous AQS protocols from the aspects of forgery and disavowal. We show that in these protocols the receiver Bob can realize existential forgery of the sender's signature under known message attack. Bob can even achieve universal forgery when the protocols are used to sign a classical message. Furthermore, the sender Alice can successfully disavow any of her signatures by simple attack. The attack strategies are described in detail and some discussions about the potential improvements of the protocols are given. Finally we also present several interesting topics in future study on AQS protocols.
\begin{description}
\item[PACS numbers]
03.67.Dd, 03.67.Ac
\end{description}
\end{abstract}

\pacs{Valid PACS appear here}
\maketitle


\section{\label{sec:level1}Introduction}

Cryptography is the approach to protect data secrecy in public environment. As we know, the security of most classical cryptosystems is based on the assumption of computational complexity and might be susceptible to the strong ability of quantum computation \cite{Shor,Grover}. Fortunately, this difficulty can be overcome by quantum cryptography \cite{GRTZ02}. Different from its classical counterpart, quantum cryptography is the combination of quantum mechanics and cryptography, where the security is assured by physical principles such as Heisenberg uncertainty principle and quantum no-cloning theorem. Now quantum cryptography has attracted a great deal of attentions because it can stand against quantum attack. Quite a few branches of quantum cryptography have been studied in recent years, including quantum key distribution (QKD) \cite{BB84,E91,B92}, quantum secret
sharing (QSS) \cite{CGL99,HBB99,KKI99}, quantum secure direct communication (QSDC) \cite{LL02,BF02,DLL03}, quantum identity authentication \cite{DHHM99,ZZ2000}, and so on.

Message authentication and digital signature are important branches of cryptography \cite{Schneier}. The former provides the ability to assure message's origin and integrity. It is used to prevent \emph{a third party} from masquerading as the legitimate users or substitute a false message for a legitimate one. The latter can provide not only the ability of message authentication, but also the function of nonrepudiation. It is used mainly to prevent the cheat from \emph{the legitimate users}, including forging the sender's signature by the receiver, and repudiating the signature by the sender.

As we know, the quantum nature makes quantum message quite different from classical one. Compared with their counterparts in classical cryptography, the authentication \cite{CSP02,BCGSA,PCS03,YL03} and signature \cite{ZK02,CL08,Z08,LCL09,ZQ10} of quantum message are more difficult. In Ref. \cite{BCGSA}, Barnum et al pointed out that if one wants to securely authenticate a quantum message he/she must do a perfect encryption on it. That is to say, anyone else can learn nothing about the content of an authenticated quantum message. Consequently, in a quantum signature protocol, which has the functions of authentication, the receiver of a signed quantum message cannot learn anything about the content. However, in an application of signature it is generally necessary for the receiver to learn something about the content of the signed message. As a result, they drew a conclusion that signing a quantum message is impossible.

Though Barnum et al's conclusion put a serious obstacle for quantum message signature, the study of quantum signature scheme has not been stopped. In 2002 Zeng and Keitel proposed a pioneering arbitrated quantum signature (AQS) protocol, which can be used to sign both classical message and quantum one \cite{ZK02}. In this protocol, the sender (signer) Alice prepares more than one copy of quantum message to be signed so that at least one copy among them exists in the signed message in the manner of plaintext. Consequently, the receiver (verifier) Bob can not only learn the content of the signed quantum message but also verify the signature with the help of the arbitrator Trent, which is not contrary to Barnum et al's conclusion. To verify the validity of a signature a necessary and important technique, i.e. probabilistic comparison of two unknown quantum states \cite{BCWW01}, is introduced in Ref. \cite{ZK02}. This work gave an elementary model to sign a quantum message, which overcomes Barnum et al's limit and is feasible in theory. In 2009 Li et al presented a Bell-states-based AQS protocol, which simplified Zeng et al's protocol by replacing Greenberger-Horne-Zeilinger states with Bell ones as the carrier \cite{LCL09}. Recently, Zou et al further simplified this protocol achieving AQS without entangled state \cite{ZQ10}. Both of them still preserve the merits in Zeng et al's protocol.

Cryptanalysis plays an important role in the development of cryptography. It estimates a protocol's security level, finds potential loopholes and tries to overcome the security issues. As pointed out by Lo and Ko, \emph{breaking cryptographic systems was as important as building them}
\cite{LK05}. In the study of quantum cryptography, quite a few effective attack strategies have been proposed, such as intercept-resend attack \cite{GGW08}, entanglement-swapping attack \cite{ZLG01,GQW07}, teleportation attack \cite{GGWcpb08}, dense-coding attack \cite{GIEEE11,HLL10,QGW06},
channel-loss attack \cite{W03,W04}, Denial-of-Service (DoS) attack \cite{C03,GGWZpra08}, Correlation-Extractability (CE) attack
\cite{GWZpla06,GLWZcpl08,GQWZoc10}, Trojan horse attack \cite{GFKZ06,DLZZ05}, participant attack \cite{GQW07,QGW06}, and so on. Understanding those attacks will be helpful for us to design new schemes with high security.

When we analyze the security of a digital signature protocol, we generally pay our attention to two important security requirements, i.e., the signature should not be forged by the attacker (including the receiver) and the signer cannot disavow his/her signature. In classical cryptography, as far as the forgery is concerned, the attacks can be classified into the following three models \cite{GMR88}.

1. \emph{Key-only attack}, where the attacker knows only the public verification key.

2. \emph{Known message attack}, where the attacker is given valid signatures for a variety of messages known by the attacker but not chosen by the attacker.

3. \emph{Adaptive chosen message attack}, where the attacker previously knows signatures on arbitrary messages of the attacker's choice.

Furthermore, the attack generally results in three kinds of results, that is,

1. \emph{Universal forgery}, which results in the ability to forge signatures for any message (also called \emph{total break} if the signing key is obtained).

2. \emph{Selective forgery}, which results in a signature on a message of the attacker's choice.

3. \emph{Existential forgery}, which results in some valid message/signature pair not already known to the attacker.

In this paper we study the cryptanalysis of AQS protocols, and focus on the forgery by the receiver Bob and the repudiation by the signer Alice. Taking protocols in Refs. \cite{LCL09,ZQ10} as examples, we will show that, in the circumstance of known message attack, Bob can give lots of existential forgeries of Alice's signature. More seriously, when the protocols are used to sign a classical message Bob can achieve universal forgery of Alice's signature. Furthermore, Alice can successfully disavow the signature she signed for Bob. Therefore, some improvements on these AQS protocols are urgently needed.

The rest of this paper is organized as follows. In Sec. II and Sec. III we respectively analyze the security of AQS protocols in Refs. \cite{LCL09} and \cite{ZQ10}, where the protocols are briefly recalled and particular attack strategies are demonstrated. Some useful discussions are given in Sec. IV, and Sec. V is our conclusion.

\section{Analysis of the AQS protocol with Bell states}

In this section we will introduce quantum one-time pad algorithm firstly, which is helpful to understand our attack strategies. Then the AQS protocol with Bell states \cite{LCL09} is described briefly and our security analysis follows.

\subsection{Quantum one-time pad}
As the analog of classical one-time pad, quantum one-time pad (QOTP), also called quantum Vernam cipher \cite{Leung02}, uses classical key bits to encrypt quantum states. This cipher plays an important role in AQS protocols and it is meaningful for us to make it clear. Boykin and Roychowdhury proved that $2n$ random classical bits are both necessary and sufficient for encrypting any unknown state of $n$ qubits in an informationally secure manner \cite{BR03}. Suppose $|P\rangle=\bigotimes_{i=1}^n|p_i\rangle$ is a quantum message composed of $n$ qubits $|p_i\rangle=\alpha_i|0\rangle+\beta_i|1\rangle$, and the key is $K\in\{0,1\}^{2n}$. The QOTP encryption $E_K$ on the quantum message can be described by
\begin{eqnarray}
|C\rangle=E_K|P\rangle=\bigotimes_{i=1}^n\sigma_x^{k^{2i}}\sigma_z^{k^{2i-1}}|p_i\rangle
\label{eq:one},
\end{eqnarray}
where $k^j$ denotes the $j$th bit of $K$, and $\sigma_x$ and $\sigma_z$ are Pauli operations. The corresponding decryption $D_K$ is
\begin{eqnarray}
D_K|C\rangle=\bigotimes_{i=1}^n\sigma_z^{k^{2i-1}}\sigma_x^{k^{2i}}|c_i\rangle
\label{eq:two},
\end{eqnarray}
where $|c_i\rangle$ denotes the $i$th qubit of the ciphertext $|C\rangle$.

\subsection{AQS protocol with Bell states}
The AQS protocol with Bell states \cite{LCL09} is as follows.

\vspace{3mm}\emph{Initializing phase.}\vspace{1mm}

Alice and Bob share a key with the arbitrator Trent, i.e. $K_A$ and $K_B$ respectively, and $n$ Bell states $|\psi_i\rangle_{AB}=\frac{1}{\sqrt{2}}(|00\rangle+|11\rangle)$ are shared between Alice and Bob.

\vspace{3mm}\emph{Signing phase.}\vspace{1mm}

S1. Alice obtains three copies of the quantum message $|P\rangle=\bigotimes_{i=1}^n|p_i\rangle$ to be signed.

S2. Using the key $K_A$, Alice encrypts one copy of $|P\rangle$ into $|R_A\rangle$ where
\begin{eqnarray}
|R_A\rangle=E'_{K_A}|P\rangle=\bigotimes_{i=1}^nM_{k_A^i}|p_i\rangle
\label{eq:three}.
\end{eqnarray}
Here $M_{k_A^i}=\sigma_x$ when $k_A^i$, the $i$th bit of $K_A$, is 0, while $M_{k_A^i}=\sigma_z$ when $k_A^i=1$.

S3. Alice performs Bell measurements on each qubit in the second copy of $|P\rangle$ and the corresponding qubit in the Bell states, obtaining the measurement result $|M_A\rangle=\bigotimes_{i=1}^n|m_A^i\rangle$, where $|m_A^i\rangle$ are random Bell states. The aim of this step is to send the second copy of message to Bob by teleportation via the Bell states previously shared between them.

S4. Alice encrypts $|M_A\rangle$ and $|R_A\rangle$ by $K_A$, obtaining the signature
$|S\rangle=E_{K_A}(|M_A\rangle\otimes|R_A\rangle)$,
where $E_{K_A}$ denotes the encryption of QOTP.

S5. Alice sends the signature and the third copy of message $|S\rangle\otimes|P\rangle$ to Bob.

\vspace{3mm}\emph{Verifying phase.}\vspace{1mm}

V1. Bob encrypts the signed message by QOTP, obtaining
$|Y_B\rangle=E_{K_B}(|S\rangle\otimes|P\rangle)$,
and sends it to Trent.

V2. Trent decrypts the received ciphertext with $K_B$ and $K_A$, obtaining $|P\rangle$, $|M_A\rangle$, and $|R_A\rangle$, and then verifies whether $|R_A\rangle=E'_{K_A}|P\rangle$ by probabilistic comparison of quantum states \cite{BCWW01}. If it is, he sets $r=1$, otherwise $r=0$.

V3. Trent recovers $|S\rangle$ and $|P\rangle$ (note that the compared states can be recovered after the comparison if they are indeed equal), reads out (and replicates) Alice's measurement result $|M_A\rangle$, and sends $|Y_{TB}\rangle=E_{K_B}(|M_A\rangle\otimes|S\rangle\otimes|P\rangle\otimes|r\rangle)$ to Bob. Here $E_{K_B}$ denotes the QOTP encryption using the key $K_B$.

V4. Bob decrypts the received ciphertext and judges whether $r=1$. If not, he believes the signature is forged and stops the protocol.

V5. According to $|M_A\rangle$, Bob can obtain the second copy of quantum message via the teleportation by Alice. Then he compares it with the copy received from Trent. Bob accepts Alice's signature when they are equal, otherwise he rejects it.

\subsection{Analysis of the AQS protocol with Bell states}

Now we analyze how the above protocol achieves the functions of digital signature. To show that we begin with the role of the arbitrator Trent. In this protocol, Trent knows $K_A$ and he can do the comparison whether $|R_A\rangle=E'_{K_A}|P\rangle$ in Step V2. When this equation holds, it implies that the signed message is really come from Alice because others do not know $K_A$. Note that after the verifying phase, all the three copies of quantum messages will be transmitted to Bob and Trent will have none of them. Furthermore, Trent does not know the content of the quantum message because he cannot read it owing to its quantum feature. Therefore, by sending his judgement result $r$ to Bob, Trent can only tell Bob whether this signed message originated from Alice. That is to say, if $r=1$, Trent ensures that Alice sent a certain quantum message (to Bob) but the content is unknown to him.

Based on the above analysis, there must be a way for Trent to resolve disputes between Alice and Bob though the protocol does not describe it clearly. Otherwise it is just like a protocol for message authentication instead of digital signature. It is not difficult to imagine the situation where dispute appears, that is, Bob says that Alice signed a message $|\mathcal{P}\rangle$ for him but Alice announces that she did not sign such a message for Bob (maybe she indeed singed a message for Bob before but it is not $|\mathcal{P}\rangle$). In this condition Trent will require Bob to provide the message $|\mathcal{P}\rangle$ and Alice's corresponding signature $|\mathcal{S}\rangle$, decrypt $|\mathcal{S}\rangle$ with $K_A$ (obtaining $|\mathcal{M}_A\rangle$ and $|\mathcal{R}_A\rangle$), and then verify whether $|\mathcal{R}_A\rangle=E'_{K_A}|\mathcal{P}\rangle$, which is just like the process in Step V2. If the comparison result is positive Trent concludes that $|\mathcal{P}\rangle$ is indeed Alice's singed message and Alice is disavowing her signature. On the contrary, Trent believes the signature is forged by Bob if the result is negative.

\subsubsection{Bob's forgery}
Let us see the possibility for Bob to forge a valid signed message of Alice first. As analyzed in Ref. \cite{LCL09}, it looks like that Bob can counterfeit Alice's signature only when he knows the key $K_A$ because in this condition he can provide $|\mathcal{P}\rangle$ and $|\mathcal{S}\rangle=E_{K_A}(|\mathcal{M}_A\rangle\otimes|\mathcal{R}_A\rangle)$ such that $|\mathcal{R}_A\rangle=E'_{K_A}|\mathcal{P}\rangle$. But $K_A$ is the key shared between Alice and Trent via QKD, which will be kept unknown to Bob. Consequently, it is impossible for Bob to forge Alice's signature in this manner. Then an interesting question arises, that is, is there other way for Bob to give a valid counterfeit of Alice's signature? Equivalently, can Bob successfully forge a signature without $K_A$? As we know, Bob, as the receiver of Alice's signature, indeed possesses Alice's valid signature of certain message. Therefore, he has the advantage to perform known message attack. In the following we will show that Bob can achieve existential forgery, where many valid message and signature pairs can be found.

According to the protocol, a valid signature of quantum message $P$ should be in the form of
\begin{eqnarray}
|S\rangle&=&E_{K_A}(|M_A\rangle\otimes|R_A\rangle)=E_{K_A}(|M_A\rangle\otimes E'_{K_A}|P\rangle)\nonumber\\
&=&E_{K_A}|M_A\rangle\otimes E_{K_A}E'_{K_A}|P\rangle
\label{eq:four}.
\end{eqnarray}
Because $E_{K_A}|M_A\rangle$ has no contributions for Trent to resolve disputes, the key point is whether Bob can find a pair of qubit sequences $(|\mathcal{P}\rangle, |\mathcal{S}'\rangle)$ which satisfies the relation
\begin{eqnarray}
|\mathcal{S}'\rangle=E_{K_A}E'_{K_A}|\mathcal{P}\rangle
\label{eq:five}.
\end{eqnarray}
Note that now Bob does not know $K_A$, but he has a valid signed message $(|P\rangle, |S\rangle)$, which implies he has a pair $(|P\rangle, |S'\rangle)$ satisfying $|S'\rangle=E_{K_A}E'_{K_A}|P\rangle$. Can Bob find a valid pair $(|\mathcal{P}\rangle, |\mathcal{S}'\rangle)$ from the known $(|P\rangle, |S'\rangle)$? The answer is yes. In fact if Bob performs one Pauli operation on each qubit in $|P\rangle$, obtaining $|\mathcal{P}\rangle$, and the same operation on the corresponding qubit in $|S'\rangle$, obtaining $|\mathcal{S}'\rangle$, the pair $(|\mathcal{P}\rangle, |\mathcal{S}'\rangle)$ will be a valid signed message.

To see it more clearly, suppose $|P\rangle=\bigotimes_{i=1}^n|p_i\rangle$. Then $|S'\rangle$ is in the form of $|S'\rangle=\bigotimes_{i=1}^n|s'_i\rangle$, where
\begin{eqnarray}
|s'_i\rangle=E_{k_A^{2i-1},k_A^{2i}}E'_{k_A^i}|p_i\rangle
\label{eq:six}.
\end{eqnarray}
When Bob performs one Pauli operation $U_i$ on every qubit pair $|p_i\rangle$ and $|s'_i\rangle$, he obtains
\begin{eqnarray}
|\mathcal{P}\rangle&=&\bigotimes_{i=1}^nU_i|p_i\rangle\label{eq:seven}\\
|\mathcal{S}'\rangle&=&\bigotimes_{i=1}^nU_iE_{k_A^{2i-1},k_A^{2i}}E'_{k_A^i}|p_i\rangle
\label{eq:eight}.
\end{eqnarray}
It is not difficult to see that $E_{k_A^{2i-1},k_A^{2i}}$ is the encryption of QOTP and $E'_{k_A^i}$ is also an encryption with Pauli operations. Therefore, the combination of these two encryptions $E_{k_A^{2i-1},k_A^{2i}}E'_{k_A^i}$ is still an encryption via one of four Pauli operations $\{I, \sigma_x, \sigma_z, \sigma_x\sigma_z\}$, where $I$ is the identity operator and $\sigma_x\sigma_z=i\sigma_y$. According to the commutative relations among Pauli operations, we have
\begin{eqnarray}
U_iE_{k_A^{2i-1},k_A^{2i}}E'_{k_A^i}=\pm E_{k_A^{2i-1},k_A^{2i}}E'_{k_A^i}U_i
\label{eq:nine},
\end{eqnarray}
and then
\begin{eqnarray}
|\mathcal{S}'\rangle=\bigotimes_{i=1}^n(\pm E_{k_A^{2i-1},k_A^{2i}}E'_{k_A^i}U_i|p_i\rangle)
\label{eq:ten}.
\end{eqnarray}
Note that every $|p_i\rangle$ is a pure state of a single particle, which is limited by the probabilistic comparison of two unknown quantum states \cite{ZQ10}. In this condition, all the minus signs in Eq.~(\ref{eq:ten}) are global phases and can be omitted. Therefore, we have
\begin{eqnarray}
|\mathcal{S}'\rangle=\bigotimes_{i=1}^nE_{k_A^{2i-1},k_A^{2i}}E'_{k_A^i}U_i|p_i\rangle=E_{K_A}E'_{K_A}|\mathcal{P}\rangle
\label{eq:eleven},
\end{eqnarray}
where Eq.~(\ref{eq:seven}) is used. Obviously, if Bob provides his counterfeit $(|\mathcal{P}\rangle, |\mathcal{S}'\rangle)$ to Trent it will always pass the verification.

So far we have found a simple way for Bob to achieve existential forgery of Alice's signature under known message attack. The attack strategy can be described as follows. Suppose Bob has a valid signed message of Alice, i.e. $(|P\rangle, |S\rangle)$, he performs $\bigotimes_{i=1}^nU_i$ ($U_i$ is any Pauli operation) on the qubits in $|P\rangle$, and the same operations on the last $n$ qubits in $|S\rangle$ (i.e. $|S'\rangle$). The resulted new pair $(|\mathcal{P}\rangle, |\mathcal{S}\rangle)$ must be a successful forgery. Because each $U_i$ can be selected from four Pauli operations at will, at least $4^n-1$ different forgeries can be found by Bob [the original one $(|P\rangle, |S\rangle)$ is not included]. Therefore, Bob can select the most preferred message $|\mathcal{P}\rangle_{pr}$ from them and say that it is the message Alice signed to him. In this condition Trent will always stand on the side of Bob though Alice is greatly aggrieved. Note that Bob can directly perform his attack when he just received Alice's signed message, or after the verifying phase, where he needs to launch the dispute and require Trent's judgement.

Finally, there is another thing which should be emphasized. As was pointed in Ref. \cite{LCL09}, the AQS protocol with Bell states can be used to sign both quantum message and classical one. It is not difficult to imagine that Bob can achieve universal forgery of Alice's signature under known message attack if the signed message is classical. For example, suppose Bob has a valid signed message of Alice, i.e. $(|P\rangle, |S\rangle)$, where $|P\rangle=\bigotimes_{i=1}^n|p_i\rangle$ is a classical message, that is, $|p_i\rangle=|0\rangle$ or $|1\rangle$. If Bob wants to forge Alice's signature on the message $|Q\rangle=\bigotimes_{i=1}^n|q_i\rangle$ ($|q_i\rangle=|0\rangle$ or $|1\rangle$), he just chooses the Pauli operations \begin{eqnarray}
\bigotimes_{i=1}^nU_i=\bigotimes_{i=1}^n\sigma_x^{p_i\oplus q_i}
\label{eq:twelve}
\end{eqnarray}
in the above attack, where $\oplus$ represents the addition module 2. In this circumstances, as a result, Bob can forge Alice's signature on any classical message he wants.

\subsubsection{Alice's disavowal}
Above we have shown that Bob can forge Alice's signature successfully. Now we consider the other security issue in quantum signature, i.e. Alice's disavowal. In fact Alice can also cheat in this AQS protocol. That is, Alice can successfully disavow any message she ever signed.

Suppose Alice signs a message (e.g. a contract) $|P\rangle=\bigotimes_{i=1}^n|p_i\rangle$ according to the steps in the protocol, and sends $(|P\rangle, |S\rangle)$ to Bob. When Trent sends $|Y_{TB}\rangle=E_{K_B}(|M_A\rangle\otimes|S\rangle\otimes|P\rangle\otimes|r\rangle)$ to Bob in Step V3, Alice modifies the states of the ciphertext corresponding to the last $n$ qubits in $|S\rangle$ (i.e.$|S'\rangle$), so that the resulted states of these qubits (denoted as $|S^A\rangle$) are not a valid signature of $|P\rangle$ any more. Note that Alice can find these qubits in the ciphertext and then disturb them while leave others unchanged because the qubit numbers in $|M_A\rangle$, $|S\rangle$, $|P\rangle$, and $|r\rangle$ are determinate, and the encryption of QOTP is qubit-by-qubit. Furthermore, Bob cannot discover Alice's modification on $|S'\rangle$ because he does not know $K_A$. Thus when Bob requires Alice to fulfil this contract at a later time, Alice can disavow this contract by announcing that it is not the one she ever signed or it was illegally modified by Bob. In this circumstances, interestingly, Trent will stand on the side of Alice.

This attack is very simple and not difficult to understand. First, the original signed message $(|P\rangle, |S\rangle)$ is really signed by Alice and then it will pass the verification of Trent ($r$=1). Second, because Alice only modified $|S\rangle$, which is a ciphertext for Bob and not useful for Bob's verification in Step V5, Bob will accept this signature without noticing Alice's attack. Third, when dispute appears Bob provides $(|P\rangle, |S^A\rangle)$ to Trent and requires his judgement. Obviously the modified signature will not pass Trent's verification and consequently Trent will agree with Alice, believing that the signature was forged by Bob.

\section{Analysis of the AQS protocol without entangled states}
In Ref. \cite{ZQ10} Zou et al improved the above AQS protocol to prevent the disavowal of Bob, and proposed a new AQS protocol without using entangled states. Here we takes the new protocol as our example to show that it is also susceptible to our attacks. Because the protocol and the attack strategies are similar with that in Sec.II we will describe them just in brief words.

\subsection{AQS protocol without using entangled states}
The AQS protocol without using entangled states \cite{ZQ10} is as follows.

\vspace{3mm}\emph{Initializing phase.}\vspace{1mm}

Three keys $K_{AB}$, $K_{AT}$, and $K_{BT}$ are shared between Alice and Bob, Alice and Trent, Bob and Trent respectively.

\vspace{3mm}\emph{Signing phase.}\vspace{1mm}

S1. Alice obtains three copies of the quantum message $|P\rangle=\bigotimes_{i=1}^n|p_i\rangle$ and encrypts each of them into $|P'\rangle$ using a random number $r$ as the key.

S2. Alice performs the following encryptions
$|R_{AB}\rangle=E_{K_{AB}}|P'\rangle$, $|S_{A}\rangle=E_{K_{AT}}|P'\rangle$, and $|S\rangle=E_{K_{AB}}(|P'\rangle, |R_{AB}\rangle, |S_{A}\rangle)$,
and sends $|S\rangle$ to Bob.

\vspace{3mm}\emph{Verifying phase.}\vspace{1mm}

V1. Bob decrypts $|S\rangle$ and sends $|Y_B\rangle=E_{K_{BT}}(|P'\rangle, |S_A\rangle)$ to Trent.

V2. Trent decrypts $|Y_B\rangle$ and verifies whether $|S_{A}\rangle=E_{K_{AT}}|P'\rangle$. He publishes $V_T=1$ and sends $|Y_B\rangle$ back to Bob if the equation holds, otherwise $V_T=0$.

V3. Bob decrypts $|Y_B\rangle$ and verifies whether $|R_{AB}\rangle=E_{K_{AB}}|P'\rangle$. If it is, he publishes $V_B=1$, otherwise $V_B=0$.

V4. When $V_T=V_B=1$, Bob accepts Alice's signature. In this condition Alice publishes $r$ and Bob recovers $|P\rangle$ from $|P'\rangle$. Finally Bob stores $(|P\rangle, |S_{A}\rangle, r)$ as the signed message.

\subsection{Analysis of the AQS protocol without using entangled states}
Compared with the one with Bell states, this protocol mainly changes in two aspects. On the one hand, the message copy for Bob is sent in the manner of QOTP encryption instead of teleportation, by which Bell states are not needed any more. On the other hand, the parameter $r$ is introduced to prevent Bob from obtaining the message content before he accepts it. Obviously, the first change has no effect on the attack strategies we proposed above. Now we analyze how the second change influences the attacks.

As far as Bob's forgery is considered, the situation is just like that in the protocol with Bell states. For example, $|S_{A}\rangle$ is also the encryption of $|P'\rangle$ by QOTP, and Trent does not know the (quantum) message content from beginning to end. Therefore, Bob can forge a signature by performing Pauli operations $\bigotimes_{i=1}^nU_i$ on the qubits in $|P'\rangle$, and the same operations on the qubits in $|S_A\rangle$. In fact, introducing the parameter $r$ brings only one difference, that is, if Bob wants to forge Alice's signature when he just received the signed message, he cannot choose suitable $\bigotimes_{i=1}^nU_i$ in order to obtain the fake message he prefers. This is because at that time the message $|P'\rangle$ is still a ciphertext encrypted by the unknown $r$. But Bob can still forge the signature after the verifying phase, where he launches the dispute and requires Trent's judgement. At that time $r$ has been published and Bob can choose suitable Pauli operations for him. As a result, Bob also achieves existential forgery of Alice's signature under known message attack. Similar to the situation in the protocol with Bell states, when the signed message is classical the forgery will become universal.

It is not difficult to see that introducing the parameter $r$ has no influence on Alice's attack, i.e. disavowal. Because Trent will sent $|S_A\rangle$ (in the form of ciphertext in $|Y_B\rangle$) back to Bob after his judgement, Alice still can disturb the states of the qubits in it so that $(|P\rangle, |S_{A}\rangle, r)$ is not a valid signed message any more. Furthermore, this attack will not be discovered by Bob because he does not know $K_{AT}$. By this way Alice can successfully disavow her signature on any message she ever signed.

\section{Discussions}
Here we analyze the reasons why our attack strategies work in AQS protocols, and try to find some ways to improve the protocols. Without loss of generality, we takes the protocol with Bell states \cite{LCL09} as example to give our analysis.

In our opinion, the following three facts are main reasons why the AQS protocol is susceptible to our attacks.

(1) Trent does not know the content of the signed message because it is quantum one. Therefore, when dispute appears Trent can only require Bob to provide the signed message $(|P\rangle, |S\rangle)$ and judges who is cheating by verifying whether Eq.~(\ref{eq:five}) holds. This fact gives the chance for Alice or Bob to change the states of $|P\rangle$ and $|S\rangle$ without being discovered.

(2) Though it can achieve high security for data encryption, QOTP is not so suitable (or enough) for AQS. On the one hand, this algorithm encrypts data qubit by qubit. Thus Alice and Bob can easily find and modify the qubits they want to change in the ciphertext, leaving the others undisturbed. On the other hand, Pauli operations are commute or anticommute with each other, which makes that $|P\rangle$ and $|S'\rangle$ still can pass Trent's verification after Bob's same Pauli operations on them. Therefore, Bob can give many existential forgeries based on one legal signed message.

(3) As the most important evidence when Trent resolves dispute, $|S'\rangle$ is the ciphertext of $|P\rangle$ by encryption with the key $K_A$, which is unknown to Bob. When Trent sends $|S'\rangle$ back to Bob, it is totally unreadable for Bob and its integrity cannot be verified. This gives Alice the chance to intercept and modify $|S'\rangle$ without being discovered, and then successfully disavow her signature later.

Based on the above analysis, the following two elementary manners can be used to improve the AQS protocol.

(1) After the verification Trent does not send $|S'\rangle$ to Bob, but stores it in his hand. When dispute appears Trent requires Bob to provide $|P\rangle$, and verifies the relation between $|P\rangle$ and the corresponding $|S'\rangle$ according to Eq.~(\ref{eq:five}). By this way both Alice and Bob have no chance to modify $|S'\rangle$ after Trent's verification. But this improvement cannot prevent Bob's forgery when he just received the signed message (i.e. before Trent's verification). Furthermore, it also has another disadvantage, that is, Trent has to store one signature (like $|S'\rangle$) once a verification happened, which greatly increases his burden.

(2) Introducing quantum message authentication into the AQS protocol to ensure the integrity of the signature $|S'\rangle$. For example, before she sends it to Bob Alice encoded $|S'\rangle$ with $K_A$ into the authenticated message $|S'_A\rangle$. Thus Trent can verify its integrity when he received $|S'_A\rangle$ from Bob. Similarly, Trent encoded $|S'_A\rangle$ with $K_B$ into the authenticated message $|S'_{AB}\rangle$ before he sends it to Bob. Thus when he received it Bob can verify whether it was modified by Alice in the transmission. As a result, the attacks from both Alice and Bob can be prevented. Nevertheless, the suitable authentication scheme still needs further study \cite{CSP02,BCGSA,PCS03,YL03}.

In addition, Hash function \cite{Schneier} is generally accepted to prevent existential forgery in classical digital signature. If we have Hash function on quantum message, it will be an effective way to stand against Bob's forgery. However, it cannot prevent Alice's disavowal, and the feasibility of such Hash function also needs further study.

\section{Conclusions}

We analyze the security of AQS protocols \cite{LCL09,ZQ10} and give attack strategies for both Alice and Bob. It is shown that Bob can achieve existential forgery of Alice's signature under known message attack. More seriously, Bob can realize universal forgery when the signed message is classical. Furthermore, Alice can disavow any of her signatures in these protocols. The strategies are demonstrated in detail and some discussions on how to improve the protocols are presented.

As we pointed in Sec.I, the AQS protocols gave an elementary model to sign a quantum message. To our knowledge, this is the only model which can overcome Barnum et al's limit \cite{BCGSA} now, and is feasible in theory. Though we find the insecurity in AQS protocols, the loopholes can be made up by the manners such as using quantum message authentication. Therefore, AQS protocols are still valuable and deserve to study further. In our opinion, the following topics are interesting and can be studied in the future. (1) Designing message authentication scheme which is suitable for AQS protocols. (2) Designing AQS protocol where the message can be signed and verified by multiparty. (3) As we know, the comparison of two unknown quantum states \cite{BCWW01} can only give probabilistic result. If Bob changes only few qubits (maybe the key qubits) in the signed message, it will not be discovered with certain probability. How to resolve this problem? (4) In a real channel there will be noises, which makes a legal signed message changes in the channel and cannot pass the verification. Can AQS protocols overcome the influence of noises? (5) The qubits in the signed message are limited to pure single-particle state in AQS protocols because the states comparison circuit will not work as expected when its inputs are two mixed states. How to realize the signature of quantum message including entangled states?

\end{document}